\documentclass[12pt,letterpaper]{JHEP3}
\pdfoutput=1
\usepackage{amsmath,graphicx,amssymb,subfigure}
\usepackage{slashed}
\usepackage{subfigure}
\usepackage{isomath}
\usepackage{amsthm}
\usepackage{enumerate}
\usepackage{stmaryrd}
\usepackage{amsfonts}
\usepackage{graphics}
\usepackage{epsfig}
\usepackage{slashed}
\usepackage{amsmath}
\usepackage{bm}

\newcommand{\be}{\begin{equation}}
\newcommand{\ee}{\end{equation}}
\newcommand{\bea}{\begin{eqnarray}}
\newcommand{\eea}{\end{eqnarray}}
\newcommand{\myvec}[1]%
{\stackrel{\raisebox{-2pt}[0pt][0pt]{\small$\rightharpoonup$}}{#1}}

\title{On the geometry outside of acoustic black holes in $2+1$-dimensional spacetime}
\author{Qing-Bing Wang$^{~1}$, Xian-Hui Ge$^{~1}$\footnote{gexh@shu.edu.cn}\\
${}^{1}$Department of Physics, College of Sciences, Shanghai University, 200444 Shanghai, China\\
}

\abstract{ Analogue black holes,  which can mimic the kinetic aspects of real black holes,  have been proposed for many years. The growth of the radial momentum toward the acoustic horizon is calculated  for acoustic black holes in flat and curved spacetimes. Surprisingly, for a freely infalling vortex approaching the acoustic black hole, the  Lyapunov exponent of the growth of the momentum at the horizon saturates the chaos bound $\Lambda_{\rm Lyapunov}\leq 2 \pi T$.   We investigate the orbits of test vortices and sound wave rays in the $2+1$-dimensional  ``curved" spacetime of an acoustic black hole. We show that the vortices orbit, the sound wave orbit, and the time delay of sound are similar to those famous effects of general relativity. These effects can be verified experimentally in future experiments.}
\keywords{Acoustic black hole, Chaos, Effective potential, Sound wave deflection, Sound wave time delay}
\begin{document}

\section{Introduction}
Analogue black holes have recently been a hot topic as they can provide interesting connections between astrophysical phenomena with tabletop experiments. Remarkably, the very recent experiments have reported that the thermal Hawking radiation and the corresponding temperature were observed in a Bose-Einstein condensate system \cite{Nova2019} (see \cite{Drori2019} for the experiment on stimulated the Hawking radiation in an optical system). In the seminal paper of Unruh \cite{Unruh}, the
idea of using hydrodynamical flows as analogous systems to mimic a few properties of black
hole physics was proposed. In this model, sound waves like light waves, cannot
escape from the horizon, and therefore it is named ¡°acoustic (sonic) black hole¡± (ABH). A moving
fluid with speed exceeding the local sound velocity through a spherical surface could in
principle form an acoustic black hole. The event horizon is located on the boundary
between subsonic and supersonic flow regions. The horizon, ergosphere, and Hawking radiation of $3+1$-dimensional static and rotating acoustic black holes were later studied in \cite{Visser}.
The spherical singular hypersurface in static superfluid was studied in \cite{Zloshchastiev}. It was shown that these shells form acoustic lenses analog to the ordinary optical lenses.

Acoustic black holes for relativistic fluids were derived from the Abelian Higgs model \cite{GeXH2019,GeXH2012,GeXH2010,GeXH2015}. Since the Abelian Higgs model describes high energy physics, the result in \cite{GeXH2010} implies that acoustic black holes might be created in high energy physical processes, such as quark matters and neutron stars.  The quasi one dimensional supersonic flow of Bose Einstein condensate (BEC) in Laval nozzle (convergent divergent nozzle) is considered in \cite{Barcel¨®} in order to find out which experimental settings can amplify the effect of acoustic Hawking radiation and provide observable signals. The acoustic black holes with non extremal black D3-brane were considered in \cite{Yu Chao}.
The particle production near the horizon of the acoustic black hole was studied by \cite{Roberto Balbinot}. The acoustic black hole geometry in viscous fluid with dissipation effect was considered in \cite{Bittencourt}.

Recently, Susskind and Zhao proposed that there is a duality between the operator growth in chaotic quantum systems, complexity and the momentum growth of a particle freely falling toward the black hole \cite{Susskind,Susskind1}.  For the Schwarzschild-AdS black hole, a test particle's momentum grows at a maximal rate \cite{Susskind} and the Lyapunov exponent obtained saturates the chaos bound $\varLambda_{\rm Lyapunov}\!\leq\!2\pi T$ proposed in \cite{chaosbound}. It seems to be a universal property because all the horizons at non-zero temperature are locally the Rindler-like.  We are going to examine the momentum growth of an infalling vortex toward acoustic black holes and check whether the chaos bound is saturated for analogue black holes.

In this paper, we explore the geodesic motion outside a $2+1$-dimensional acoustic black hole. We will concentrate on predicting the orbits of ``test" particles (i.e. vortices) in fluids and sound rays in the ``curved" acoustic spacetime.  We will exhibit the Lyapunov exponent at the event horizon and some well-known effects analogous to those in general relativity-- the vortices orbits, the sound wave ray deflection, the time delay of sound waves for acoustic black holes. These results can be verified by two-dimensional spherically symmetric superfluid experiments.

The paper is organized as follows: In section 2, we briefly review the geometry of acoustic black holes. In section 3, considering an infalling vortex towards acoustic black holes, we calculate the growth of the radial momentum. In section 4, we further consider the Lyapunov exponent of radial motion of vortices toward the event horizon of the acoustic black hole which is embedded in the Schwarzschild-AdS. We then study the orbit of vortices and the corresponding stability in section 5. The sound wave trajectory is then studied in section 6 in which the deflection and the time delay of the sound wave prorogation are studied. The conclusion and discussions are provided in the last section.

\section{The metric of acoustic black holes in flat and curved spacetimes}
The metric of acoustic black holes has been studied in many papers. In this paper, we first consider the general acoustic black hole metric given by Unruh, which is derived from the fluid continuity equation in a uniform fluid medium \cite{Unruh,Visser}
\bea\label{metric}
ds^2&=&-g_{tt}dt^2+g_{rr}dr^2+g_{\theta\theta}d\theta^2+g_{\phi\phi}d\phi^2\nonumber\\&=&\!-\frac{\rho_0}{c_s}\left((c_s^2-v^2)dt^2+\frac{c_s^2}{c_s^2-v^2}dr^2+r^2(d\theta^2+\sin^{2}\theta d\phi^2)\right)\!.
\eea
Here $c_s$ refers to the speed of sound in the fluid medium, and $\rho_0$ is the fluid density. The speed $v$ is the flow rate of the fluid. Assuming $\rho_0$ and $c_s$ are two constants and coordinates independent, absorbing the factor $\frac{\rho_0}{c_s}$ into the line element on the left and setting $ c_s=1 $, we can write
\be
ds^2=\!-(1-v^2)dt^2+(1-v^2)^{-1} dr^2+r^2(d\theta^2+\sin ^2\theta d\phi^2).\nonumber
\ee
The metric written in a matrix form is
\be \label{flat}
g_{\mu\nu}=
\begin{pmatrix}-(1-v^2)&0&0&0\\0&(1-v^2)^{-1}&0&0\\0&0&r^2&0\\0&0&0&r^2 \sin^2\theta
\end{pmatrix}\!\!.
\ee
\begin{figure}[htbp]
	\centering
	\includegraphics[width=0.6\linewidth,
	trim=60 80 40 20,
	clip]{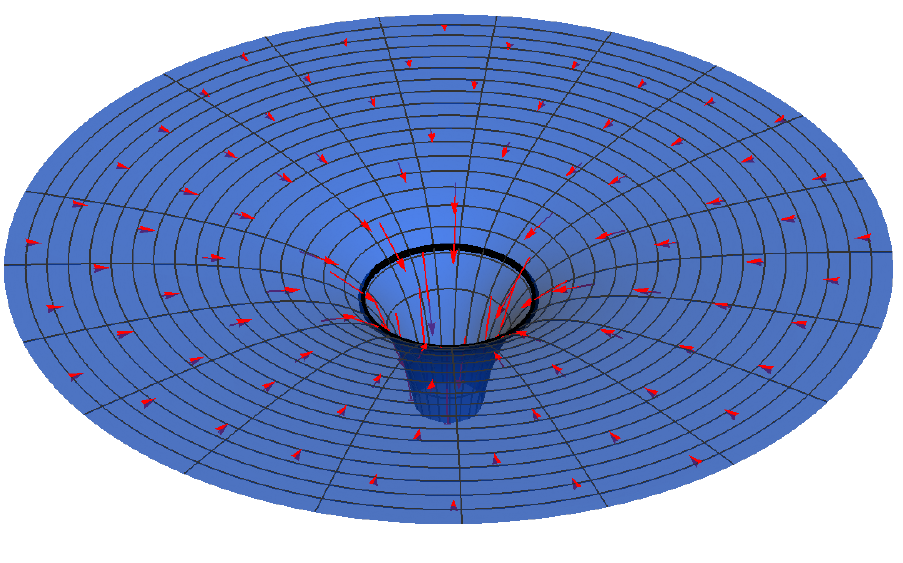}
	\caption{Schematic view of a two-dimensional acoustic black hole. The arrow represents the direction of fluid flow. The black circle is the acoustic horizon where the fluid velocity reaches the sound velocity. The central region in the ring represents the acoustic black hole. For the region in the central black circle, we can regard it as a sink leading to the high-dimensional space from which the fluid flows to the third dimensional space. }
	\label{fig:imgsoundbh}
\end{figure}

For a spatially two-dimensional fluid model, we can take $\theta=\frac{\pi}{2}$, so that $d\theta^2=0,\sin ^2\theta=1$. It is also known from fluid continuity and incompressibility $\boldsymbol{\triangledown}\!\cdot\!\boldsymbol{v}=0$ (for outside the central area), the speed $v$ should satisfy $v=-\lambda/r$, where $\lambda$ is a positive constant  and the negative sign means that the direction of the fluid velocity points to the center of the acoustic black hole (see Fig.\ref{fig:imgsoundbh}). Then the 2+1-dimensional acoustic black hole metric can be written as \cite{Susskind,Jia-Rui Sun}
\bea\label{metric1}
ds^2&=&\!-f(r)dt^2+\frac{1}{f(r)}dr^2+r^2 d\phi^2\!,\nonumber\\
f(r)&=&\!1-\frac{\lambda^2}{r^2}.
\eea
The acoustic event horizon locates at $r_0\!=\!\lambda$. The Hawking temperature of the acoustic black hole is given by \cite{Unruh,Visser,Rezn¨ªk}
\be
T=\frac{f'(r)|_{r=\lambda}}{4\pi}=\frac{1}{2\pi\lambda}.
\ee

Acoustic black holes can be embedded in curved spacetime and results a special metric. This is because in curved spacetime, in addition to accretion disks, there may be some quantum superfluids around massive bodies and black holes \cite{GeXH2019,Berezhiani}.
In the literature \cite{GeXH2019}, the authors  considered the cosmic microwave surrounded astrophysical black holes, and obtained the metric of a black hole in the curved spacetime background from the perspective of fluid mechanics
\bea
ds^2&=&(g^{GR}_{\mu\nu}\ast g^{ABH}_{\mu\nu})dx^{\mu}dx^{\nu}\nonumber\\
&=&\frac{c_s}{\sqrt{c^2_s-v_{\mu}v^{\mu}}}\bigg[(c^2_s-v_rv^r)g^{GR}_{tt}
d\tilde{\tau}^2+c^2_s \frac{c^2_s-v_{\mu}v^{\mu}}{c^2_s-v_rv^r}g^{GR}_{rr}dr^2\nonumber\\
&+&(c^2_s-v_{\mu}v^{\mu})g^{GR}_{\vartheta\vartheta}d\vartheta^2+g^{GR}_{\phi\phi}(c^2_s-v_{\mu}v^{\mu})d\phi^2\bigg].
\eea
Intriguingly, this is an analogue black hole embedded in the spacetime governed by general relativity. Among them, $g^{GR}_{\mu\nu}$ is the spacetime metric of gravity and $g^{ABH}_{\mu\nu}$ describes the geometry of the acoustic black hole, where $\mu,\nu=0,1,2,3$. The quantity $c_s$ and $v_\mu$ are the sound velocity and the fluid four-velocity respectively. For a black hole in the background of thermal radiation, the event horizon is at $r_0=2GM/c^2$. At $r_s = 6GM/c^2$, the acoustic velocity of thermal radiation is $v=c/\sqrt{3}$ with $c$ the speed of light. In the region of $r<r_s$, the thermal radiation sound wave cannot escape, that is, $r_s$ corresponds to the acoustic horizon.

In this paper, we consider some kinds of fluid surrounded black holes in AdS space. The spacetime metric of AdS-Schwarzschild black holes is \cite{Hartnoll}
\bea\label{metric2}
ds^2_{GR}&=&-r^2\left(1-\frac{r_0^2}{r^2}\right)g_{tt}+\frac{1}{r^2\left(1-\frac{r_0^2}{r^2}\right)}g_{rr}+r^2d\phi^2,
\eea
where $r_0$ is the radius of the black hole horizon, and we take AdS radius to be 1 and $G=M=c=1$. Combined with the acoustic metric given in equation (\ref{metric}), we write down the acoustic black hole metric embedded in AdS-Schwarzschild spacetime as follows
\bea\label{metric3}
ds^2&=&(g^{GR}_{\mu\nu}\ast g^{ABH}_{\mu\nu})dx^{\mu}dx^{\nu}=
\mathcal{G}_{tt}dt^2+\mathcal{G}_{rr}dr^2+\mathcal{G}_{\phi \phi}d\phi^2.
\eea
By setting  $c^2=1$ and $c_s^2=1/3$ and taking the sign of $g^{GR}_{\mu\nu}$ as $\eta_{\mu\nu}$ and the sign of $g^{ABH}_{\mu\nu}$ as $\delta_{\mu\nu}$, we can write the metric components as
\bea
&&\mathcal{G}_{tt}=-\frac{1}{3}f_{ABH}(r)f_{GR}(r),\,~ \mathcal{G}_{rr}=\frac{1}{f_{ABH}(r)f_{GR}(r)},\,~ \mathcal{G}_{\phi \phi}=r^4.\nonumber\\
&&f_{ABH}(r)=1-\frac{3\lambda^2}{r^2},\,~~~~~
f_{GR}(r)=r^2\left(1-\frac{r_0^2}{r^2}\right).
\eea
Note that the horizon $\sqrt{3}\lambda$ of the acoustic black hole is required to be larger than that event horizon $r_0$ of the black hole. Inside the event horizon that $\mathcal{G}_{tt}$ is less than 0 and $\mathcal{G}_{rr}$ is greater than 0; between the two horizons, i.e. $\sqrt{3}\lambda > r > r_0$, the opposite is true, $\mathcal{G}_{tt}>0$ and $\mathcal{G}_{rr}<0$.

In a curved spacetime, the Hawking temperature at the acoustic horizon is \cite{GeXH2019}
\be
T=\frac{1}{4\pi\sqrt{\mathcal{G}_{rr}}}\bigg(-\sqrt{\frac{g^{ABH}_{tt}}{-g^{GR}_{tt}}}g'^{GR}_{tt}+\sqrt{\frac{-g^{GR}_{tt}}{g^{ABH}_{tt}}}g'^{ABH}_{tt}\bigg)\bigg|_{r=\sqrt{3}\lambda}=\frac{3\lambda^2-r_0^2}{6\pi\lambda},
\ee
where the prime represents derivative with respect to $r$. If $g^{ABH}_{\mu\nu}$ and $c_s$ are taken as $\delta_{\mu\nu}$ and 1 respectively, the metric (\ref{metric3}) reduces to the form (\ref{metric2}) of the AdS-Schwarzschild spacetime. On the other hand, if we take $g^{GR}_{\mu\nu}$ and $c_s$ to be $\eta_{\mu\nu}$ and 1 respectively, then the metric returns to the metric given in (\ref{metric1}).

\section{Falling into the $(2+1)$-dimensional acoustic black holes: vortex motion and chaos}

In \cite{Susskind}, it was conjectured that as an object falls toward a black hole, the increase of bulk radial momentum is related to the growth of boundary operators. In this section, we study the momentum growth of an infalling vortex toward an acoustic black hole. Unstable orbits and the momentum growth can be quantified by their Lyapunov exponents \cite{Cornish}. We use Lyapunov exponent to describe the motion of a vortex near the acoustic black hole. It was proposed that vortices can behave as relativistic particles with their dynamics governed by the fluid metric \cite{Volovik03,zhang04} and their stability ensured by a topological number. Vortices with mass $m_0$ given by the Einstein's relation $E=m_0c^2$ \cite{popov73,duan94} cannot propagate at velocities faster than the sound speed. Let us consider a vortex with mass $m_0=1$ freely falling into the acoustic black hole along the radial axis. The acoustic metric is equation (\ref{metric1}). The action of the geodesic motion of the infalling vortex is \cite{Zhenhua Zhou}
\begin{equation}
	S=-\int\sqrt{-g_{\alpha\beta}u^\alpha u^\beta}d\tau,
\end{equation}
where $u^\alpha=dx^\alpha/d\tau$ is tangent to the world line and $x^\alpha$ is the spacetime coordinate, $\tau$ is an arbitrary parameter of the vortex world line. The equation of motion from the above action is
\begin{equation}\label{Vmac31}
	\dot{u}^\mu+\Gamma^\mu _{\alpha\beta}u^\alpha u^\beta-\frac{\dot{\eta}u^\mu}{2\eta}=0.
\end{equation}
where $\eta=-g_{\mu\nu}u^\mu u^\nu>0$ and the dot represents the derivative with respect to $\tau$. Note that the acoustic black hole metric in general does not satisfy the Einstein equation. The canonical momentum can be obtained by
\begin{equation}\label{Vmac32}
	p_\alpha=\frac{\delta S}{\delta u^\alpha}=\frac{g_{\alpha\beta}u^\beta}{\sqrt{-g_{\mu\nu}u^\mu u^\nu}}.
\end{equation}
We choose the gauge $\tau=t$ and take the ansatz $r=r(t)$, $\phi=$ constant. Then, equation (\ref{Vmac31}) and equation (\ref{Vmac32}) reduce to
\bea\label{Vmac33}
\dot{r}&=&-\left(-g^{rr}g_{tt}-\frac{g^{rr}g^2_{tt}}{A^2}\right)^{1/2},
\\
p_r&=&\left(-A^2 g^{tt}g_{rr}-g_{rr}\right)^{1/2},\label{Vmac331}
\eea
where $\dot{r}<0$ means the vortex falls into the acoustic black hole and $A=\sqrt{g^{2}_{tt}/\eta}>0$ is an integral constant. We substitute equation (\ref{metric1}) into (\ref{Vmac33}) and obtain
\bea\label{Vmac34}
\dot{r}&=&-\left(f^2 - \frac{f^3}{A^2}\right)^{1/2},\\
p_r&=&\left(A^2 f^{-2} - f^{-1}\right)^{1/2}.\label{Vmac341}
\eea
We evaluate the growth rate of the momentum in the Rindler coordinate \cite{A. R. Brown,Sang Pyo Kim}. Since $d\rho/dr\sim 1/\sqrt{f}$, the Rindler momentum $p_\rho$ near the acoustic horizon is given by
\begin{equation}\label{Vmac35}
	p_\rho\sim \sqrt{f}p_r\sim \left(A^2 f^{-1}-1\right)^{1/2}.
\end{equation}
We can find $(r-\lambda)\propto e^{-4\pi Tt}$ in equation (\ref{Vmac34}) near the acoustic black hole horizon, where $T=1/(2\pi\lambda)$ is the Hawking temperature of the acoustic black hole. Near the acoustic horizon, the radial momentum $p_\rho$ can be approximated as
\begin{equation}\label{Vmac36}
	p_\rho\propto(r-\lambda)^{-1/2}\sim e^{2\pi Tt}.
\end{equation}
We then obtain $p_\rho\sim e^{2\pi Tt}$. The growth of the radial momentum near the acoustic horizon is described by the Lyapunov exponent $\varLambda_{\rm Lyapunov}$
\begin{equation}\label{Vmac37}
	\varLambda_{\rm Lyapunov}=2\pi T.
\end{equation}
 It is interesting to note that the chaos bound is satisfied by the momentum growth of the ``test" particle in the geometry of acoustic black holes. This shows that acoustic black holes are similar to the real black hole in dynamics, which provides the possibility for the experimental simulation of the chaotic behavior near the black hole horizon.

\section{Vortex motion and chaos for acoustic black holes in curved spacetime}

In order to examine the universality of the relation $	\varLambda_{\rm Lyapunov}=2\pi T$, we consider a special situation in which the acoustic black hole is embedded in a Schwarzschild-AdS spacetime. Through the Hadamard product, we consider that the metric (\ref{metric3}) of the acoustic black hole in the curved space-time of $(2+1)$-dimensional AdS-Schwarzschild.

Similarly, the action in curved spacetime is considered as
\be
	S=-\int\sqrt{-\mathcal {G }_{\alpha\beta}u^\alpha u^\beta}d\tau.
\ee
We can obtain the radial momentum
\bea
\dot{r}&=&\left(\frac{f_{ABH}^{2}f_{GR}^{2}}{3}-\frac{f_{ABH}^{3}f_{GR}^{3}}{9 \mathcal{A}^2}\right)^{1/2},
\label{Vmac41}\\
p_r&=&-\mathcal{A}\left(3 f_{ABH}^{-2}f_{GR}^{-2} - \frac{f_{ABH}^{-1}f_{GR}^{-1}}{\mathcal{A}^2}\right)^{1/2},\label{Vmac411}
\eea
where $\mathcal{A}=\sqrt{\mathcal{G}^{2}_{tt}/\eta_1}>0, $ is the integral constant with the timelike condition $\eta_1=-\mathcal{G}_{\mu\nu}u^\mu u^\nu>0$.

Since $d\rho/dr=1/\sqrt{f_{ABH}f_{GR}}$,  the Rindler momentum $p_\rho$ is
\be
p_\rho=\left(\frac{3\mathcal{A}^2}{f_{ABH}f_{GR}}-1\right)^{1/2}.
\ee
Near the acoustic horizon, the radial momentum $p_\rho$ can be approximated as
\be
p_\rho\approx\left(\frac{3\sqrt{3}\mathcal{A}^2}{2(3\lambda-\sqrt{3}r_0)(r-\sqrt{3}\lambda)}\right)^{1/2}\sim(r-\sqrt{3}\lambda)^{-1/2}.
\ee
By calculating the expression (\ref{Vmac41}), we can obtain $(r-\lambda)\propto e^{-4\pi Tt}$, where $T=\frac{3\lambda^2-r_0^2}{6\pi\lambda}$. We can find that the Lyapunov exponent $\varLambda_{\rm Lyapunov}$ of the stable acoustic black hole in AdS-Schwarzschild space-time (i.e. the temperature $T$ is a constant at the acoustic horizon) is equal to ${2\pi T}$ and still satisfies the relation $\varLambda_{\rm Lyapunov}=2\pi T$. This is consistent with the results of pure acoustic black holes obtained in equation (\ref{Vmac37}). It shows that for different types of acoustic black holes, the increase of radial momentum and the growth of operator size still satisfy the chaotic boundary conditions.

Interestingly, we found that $p_\rho\sim T^{-1/2}e^{2\pi Tt}$ near the  acoustic horizon.
Moreover, we find that there is a similar law in the pure acoustic black hole or optical black hole, that is, the momentum growth of particles or vortices near the horizon have a $T^{-1/2}$ factor besides the e-exponential relationship with the Hawking temperature, which indicates that the Hawking temperature has a promoting or inhibiting effect on the exponential growth of $p_\rho$.
At low temperature, the growth of radial momentum $p_\rho$ is  faster by a factor $T^{-1/2}$ than just exponential $\varLambda_{\rm Lyapunov}$.
\section{Vortices orbits}

In the above two sections, we consider the momentum growth of the vortex as the test particle near the horizon of the acoustic black hole. In this section, we will study the trajectory of vortices in acoustic black holes. We will consider the falling trajectory of vortices with different initial energy and angular momentum, and numerically calculate the increase of radial momentum corresponding to the growth of operator size.

\subsection{Effective potential}
Since the metric is independent of $t$ and $\phi$, the quantities $\bm{\xi} \cdot \mathbf{u}$ and $\bm{\eta} \cdot \mathbf{u}$ are conserved, where $\mathbf{u}$ is the four-velocity of the vortex, $\xi^\alpha=(1,0,0,0)$ and $\eta^\alpha=(0,0,0,1)$ are killing vectors. The conserved energy per unit of static mass is \cite{Hartle,Carroll}
\begin{equation}\label{key0}
\epsilon=-\bm{\xi} \cdot \mathbf{u}=(1-v^2)\frac{dt}{d\tau}.
\end{equation}
The conservation angular momentum per unit of rest mass is
\begin{equation}\label{key00}
l=\bm{\eta} \cdot \mathbf{u}=r^2\sin^2\theta\frac{d\phi}{d\tau}.
\end{equation}
The four-velocity vectors satisfy the relation
\be\label{ep1}
\mathbf{u} \cdot \mathbf{u}=g_{\alpha \beta}u^\alpha u^\beta = -1.
\ee
Substituting \eqref{metric} into \eqref{ep1} and taking into account the equatorial plane condition $u^\theta=0$, $\theta=\pi/2$, we arrive at
\begin{equation}\label{key01}
-(1-v^2)(u^t)^2+(1-v^2)^{-1}(u^r)^2+r^2(u^\phi)^2=-1.
\end{equation}
Writing $u^t=dt/d\tau$, $u^r=dr/d\tau$, and $u^\phi=d\phi/d\tau$, and substituting \eqref{key0} and \eqref{key00} into \eqref{key01} to eliminate $dt/d\tau$ and $d\phi/d\tau$, we obtain
\be
-(1-v^2)^{-1}\epsilon^2+(1-v^2)^{-1}\bigg(\frac{dr}{d\tau}\bigg)^2+\frac{l^2}{r^2}=-1.
\ee
The square of the conservation energy per unit of rest mass can be obtained as
\be
\epsilon^2=\bigg(\frac{dr}{d\tau}\bigg)^2+\left(\frac{l^2}{r^2}+1\right)(1-v^2).\\\label{key000}
\ee
By defining the effective energy \cite{Carter,Bardeen,Felice}
\begin{equation}\label{key1}
\text{\Large$\varepsilon$}\equiv\frac{\epsilon^2-1}{2}=\frac{1}{2}\bigg(\frac{dr}{d\tau}\bigg)^2+V_{\mathrm{eff}}(r),
\end{equation}
where the effective potential is
\be
V_{\mathrm{eff}}(r)=\frac{1}{2}\bigg[\bigg(\frac{l^2}{r^2}+1\bigg)(1-v^2)-1\bigg]=\frac{1}{2}\bigg(-v^2+\frac{l^2}{r^2}-v^2\frac{l^2}{r^2}\bigg),\label{key11}
\ee
\begin{figure}[htbp]
	\centering
	\includegraphics[width=0.7\linewidth]{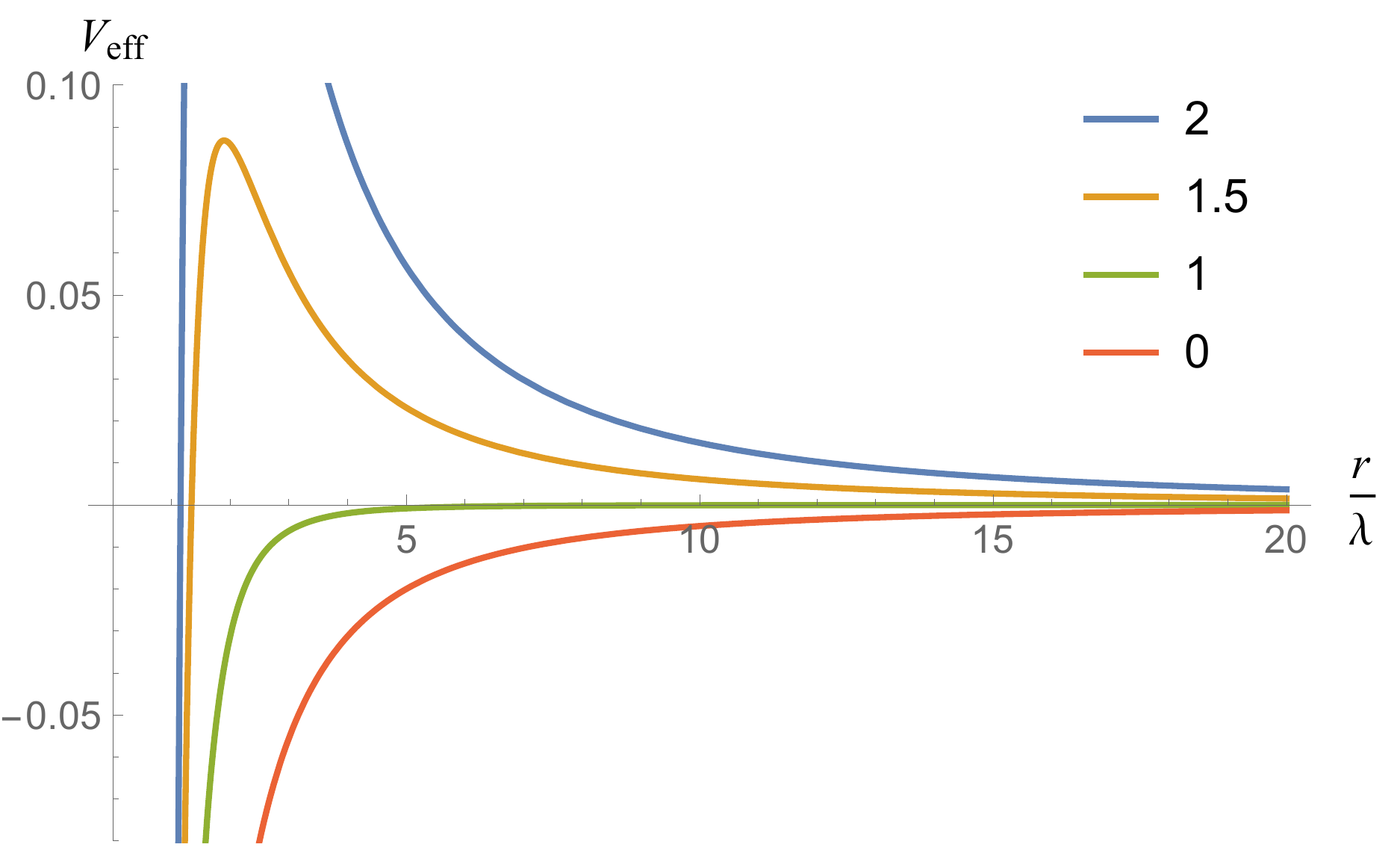}
	\caption{
		We plot the effective potential $V_{\mathrm{eff}}$ as a function of $r/\lambda $ by setting $l/\lambda=0, 1, 1.5, 2$. It can be seen from the figure that there is no minimum point for the effective potential, that is, there is no stable circular orbit for the vortex.}
	\label{fig:imgveff}
\end{figure}
and $v=-\lambda/r$, we obtain the expression of the effective potential
\be
V_{\mathrm{eff}}(r)=\frac{1}{2}\bigg[\bigg(\frac{l^2}{r^2}+1\bigg)(1-\frac{\lambda^2}{r^2})-1\bigg]=\frac{1}{2r^4}[(l^2-\lambda^2)r^2-\lambda^2l^2].\label{effp1}
\ee
For a concrete acoustic black hole, we know that $\lambda$ is a positive constant, and $l^2\geq 0$.
If the first order derivative of the effective potential $V_{\mathrm{eff}}'(r) =0$, the critical radius satisfies $r_\ast^2=\frac{2\lambda^2l^2}{l^2-\lambda^2}$. When $l^2>\lambda^2$, we can obtain $r_\ast=\frac{\sqrt{2}\lambda l}{\sqrt{l^2-\lambda^2}}$ and the second derivative of the effective potential $V_{\mathrm{eff}}''(r_\ast)=-\frac{4 \lambda^2l^2}{r_\ast^6}<0$, which shows  that $r_\ast$ is the radius of the unstable orbit. When $l^2\leq\lambda^2$, we can find $V_{\mathrm{eff}}'(r)>0$ and $V_{\mathrm{eff}}''(r)<0$, which means that the effective potential $V_{\mathrm{eff}}(r)$ increases with the increase of $r$  and there is no stable orbit \cite{Rezzolla,MTW}.

\begin{figure}[htbp]
	\centering
	\includegraphics[width=0.8\linewidth]{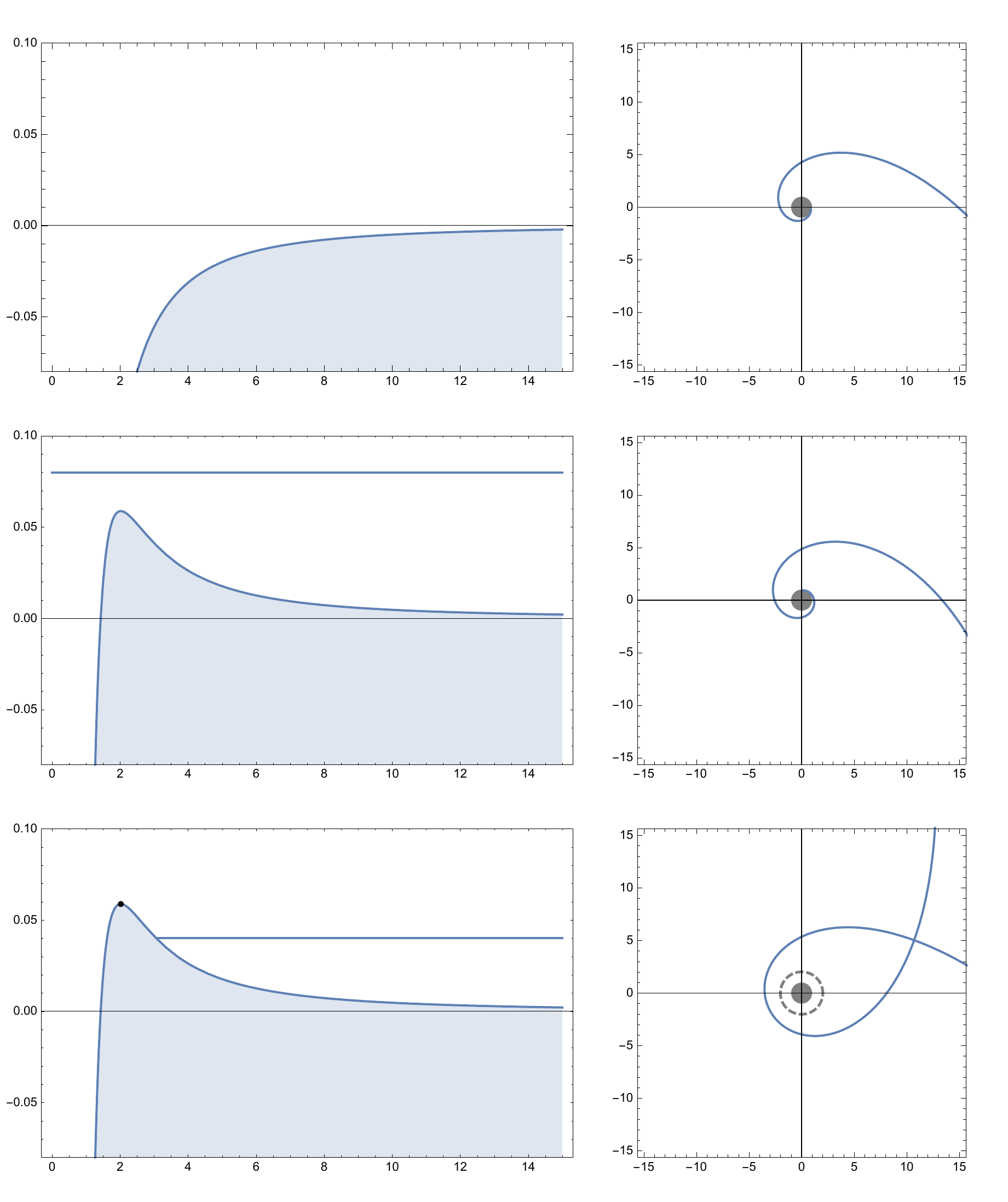}
	\caption[Potential function and vortex orbits] {From the first row, it can be seen that when $l=0$, the potential function increases monotonically with $r$, and the vortex always falls straight into the  acoustic black hole. The case of $l^2<\lambda^2$ is similar. The value $l/\lambda$ is $1.4$ in other parts. For $l^2>\lambda^2$, if the vortex energy is large enough, it will fall into the  acoustic black hole, as shown in the second row. If the vortex energy is low, when the vortex is near the acoustic black hole, it will escape from the acoustic black hole, as shown in the third pair of pictures, where the dotted line on the right is the position of unstable orbit, corresponding to the maximum potential energy in the left figure. }
	\label{fig:imgorbits1}
\end{figure}
As shown in Fig.\ref{fig:imgveff}, the effective potential energy $V_{\mathrm{eff}}(r)$ does not have a minimum value, so there is no stable circular orbit for the acoustic black hole of this type.

\subsection{Radial infalling orbit}
From the effective energy equation (\ref{key1}) and the effective potential equation (\ref{key11}), we can study the radial motion of the vortices. From the equation (\ref{key1}) with $\epsilon=1$ and $l=0$ , there is
\begin{equation}\label{key3}
0=\frac{1}{2}\bigg(\frac{dr}{d\tau}\bigg)^2-\frac{1}{2}v^2=\frac{1}{2}\bigg(\frac{dr}{d\tau}\bigg)^2-\frac{1}{2}\frac{\lambda^2}{r^2},
\end{equation}
where $r$ decreases with the increase of time in which the four-velocity component is given by $dr/d\tau=v<0$. Taken together with the time component $dt/d\tau$ given by equation (\ref{key0}), the four-velocity is
\be
u^\alpha=((1-v^2)^{-1},v,0,0)=\bigg(\bigg(1-\frac{\lambda^2}{r^2}\bigg)^{-1},-\frac{\lambda}{r},0,0\bigg).
\ee
Considering that $\lambda>0$, we take a negative sign. Rewritten (\ref{key3}) in the form
\be
\frac{1}{v}dr=d\tau,
\ee
\be
-\bigg(\frac{\lambda^2}{r^2}\bigg)^{-\frac{1}{2}}dr=-\frac{r}{\lambda}dr=d\tau,
\ee
the integral result is
\be\label{rfo1}
r(\tau)=(-2\lambda(\tau-\tau_*))^{1/2}=(2\lambda)^{1/2}(\tau_*-\tau)^{1/2},
\ee	
where $\tau_*$ is an integral constant. Computing $dt/dr$ from equation (\ref{key0}) with $\epsilon=1$ and equation (\ref{key3}), we obtain
\be
\frac{dt}{dr}=(1-v^2)^{-1}v^{-1}=-\bigg(1-\frac{\lambda^2}{r^2}\bigg)^{-1}\frac{r}{\lambda}=-\frac{r^3}{\lambda(r^2-\lambda^2)}.
\ee
After integration, we have
\be\label{rfo2}
t=\int \!(1-v^2)^{-1}v^{-1}dr =t_*-\frac{r^2+\lambda^2\ln(r^2-\lambda^2)}{2\lambda},
\ee
where $t_*$ is another integral constant. From \eqref{rfo2}, when $r\rightarrow\infty,t\rightarrow-\infty$, the vortex is falling from infinity. From equation \eqref{rfo1} we can see that for any fixed $r_1$ value outside the horizon, it takes only a finite amount of proper time to reach $r_2=\lambda$, while equation \eqref{rfo2} shows that it takes an infinite quantity of coordinate time $t$. This is just a sign that the coordinates are flawed at $r=\lambda$.
This shows that for the falling object A itself, the time from $r_1$ to $r_2$ is limited. For the observer B with fixed coordinates, he observed that A keeps approaching the horizon $r_2$. The observation here refers to the sound waves received by B from A. Replacing $\omega_{\infty}$, $v_{\infty}$ with $\omega_{r_1}$, $v_{r_1}$ and replacing $\omega_{*}$, $v_{*}$ with $\omega_{r_2}$, $v_{r_2}$ in \eqref{ars2}, we then obtain $\omega_{r_1}=0$. The sound waves from $r_2$ are infinitely redshift \footnote{For the discussions on redshift of sound wave propagation, one may refer to the appendix.}. B receives A's sound frequency is becoming lower and lower, and finally A's voice seems to solidify at $r_2$.

\begin{figure}[htbp]
	\centering
	\includegraphics[width=0.6\linewidth]{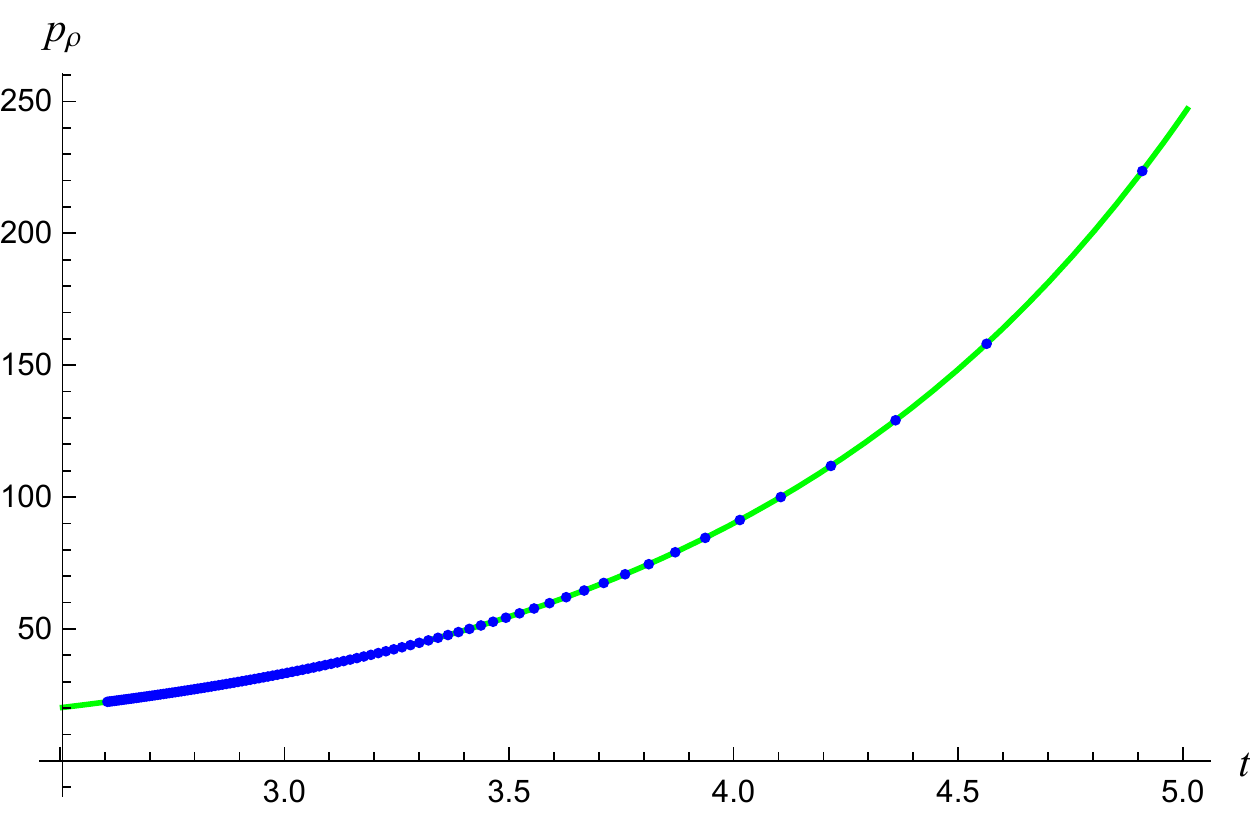}
	\caption{In the figure, the blue points are the values of $p_\rho$ and $t$ calculated by numerical solution, and the green line is the curve of exponential fitting of data. Setting $t_*=0$, $\lambda=1$, which means $2\pi T=1$, and $r$ near the horizon, the fitting result is $p_\rho\!\sim\! e^{0.9997t}$.}
	\label{fig:imgprout}
\end{figure}
We can also use formula \eqref{rfo2} to verify the relationship between the radial momentum of vortex falling and time. From equations \eqref{Vmac33}, \eqref{Vmac331} and \eqref{Vmac35}, we get $p_\rho\!\sim\! -\dot{r}/(\sqrt{f \eta})$. Further considered the formula \eqref{rfo2}, the relationship between $p_\rho$ and $t$ can be obtained. Setting $t_*=0$, $\lambda=1$, thus $2\pi T=1$, and $r$ near the horizon, we get $p_\rho\!\sim\! e^{0.9997t}$ as shown in figure \ref{fig:imgprout}. Notice that $t_*$ only shifts the $p_\rho\!-\!t$ curve horizontally, it doesnot affect the exponential relationship between $p_\rho\!-\!t$. After careful verification, we observe that as $r$ approaches the horizon $\lambda$, $(\ln p_\rho)/t$ approaches $2\pi T$. This shows that the acoustic black hole still satisfies the boundary conditions of operator scale and momentum growth, which provides a theoretical basis for simulating the chaotic behavior of particles falling into the real black hole.

\section{Sound wave orbits}
In addition to the computation of the Lyapunov exponent, the stability of the sound wave orbits and the Shapiro time delay of sound propagation near acoustic black hole deserve further investigations. In what follows, we first calculate the deflection angle and then  the time delay of sound wave propagation in $(2+1)$-dimensional acoustic black hole background in flat spacetime. 
\subsection{Sound wave deflection}
\begin{figure}[htbp]
	\centering
	\includegraphics[width=0.5\linewidth]{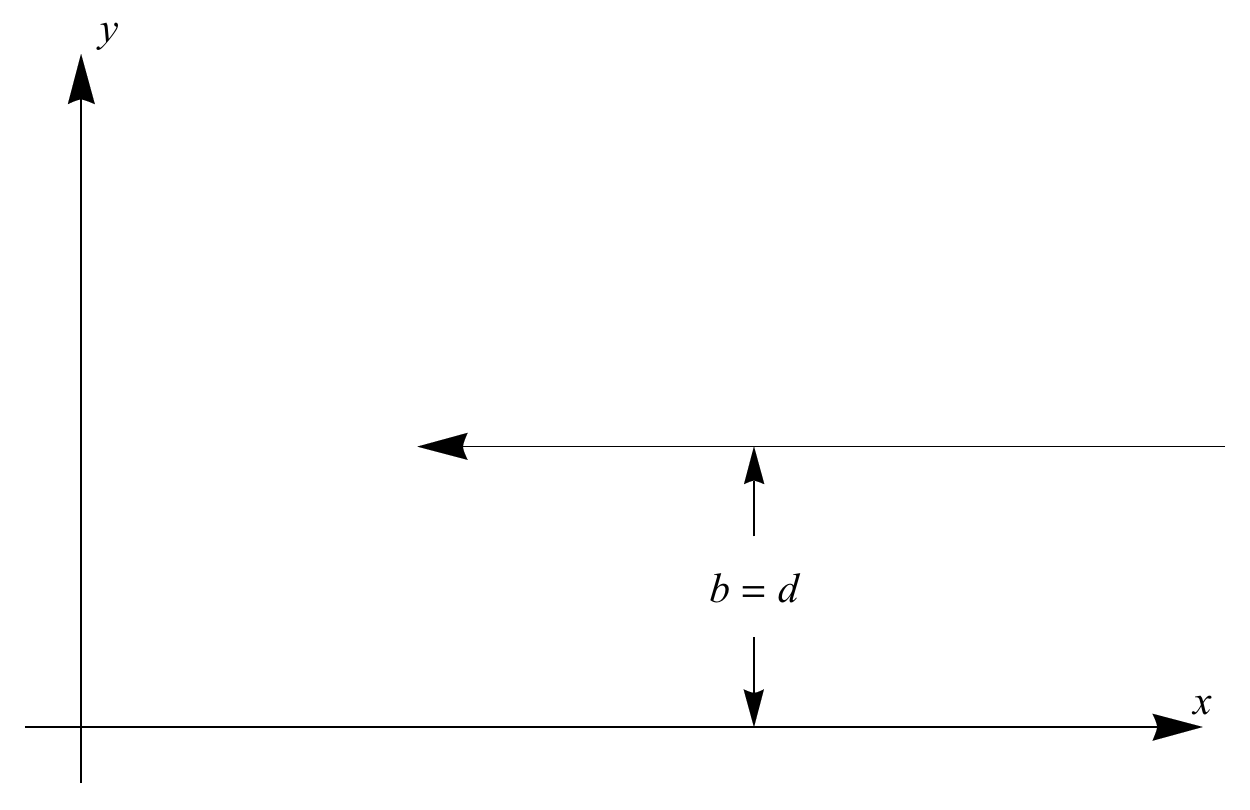}
	\caption{The orbit of the sound wave comes from infinity}
	\label{fig:imgdistance}
\end{figure}

\begin{figure}[htbp]
	\centering
	\includegraphics[width=0.8\linewidth]{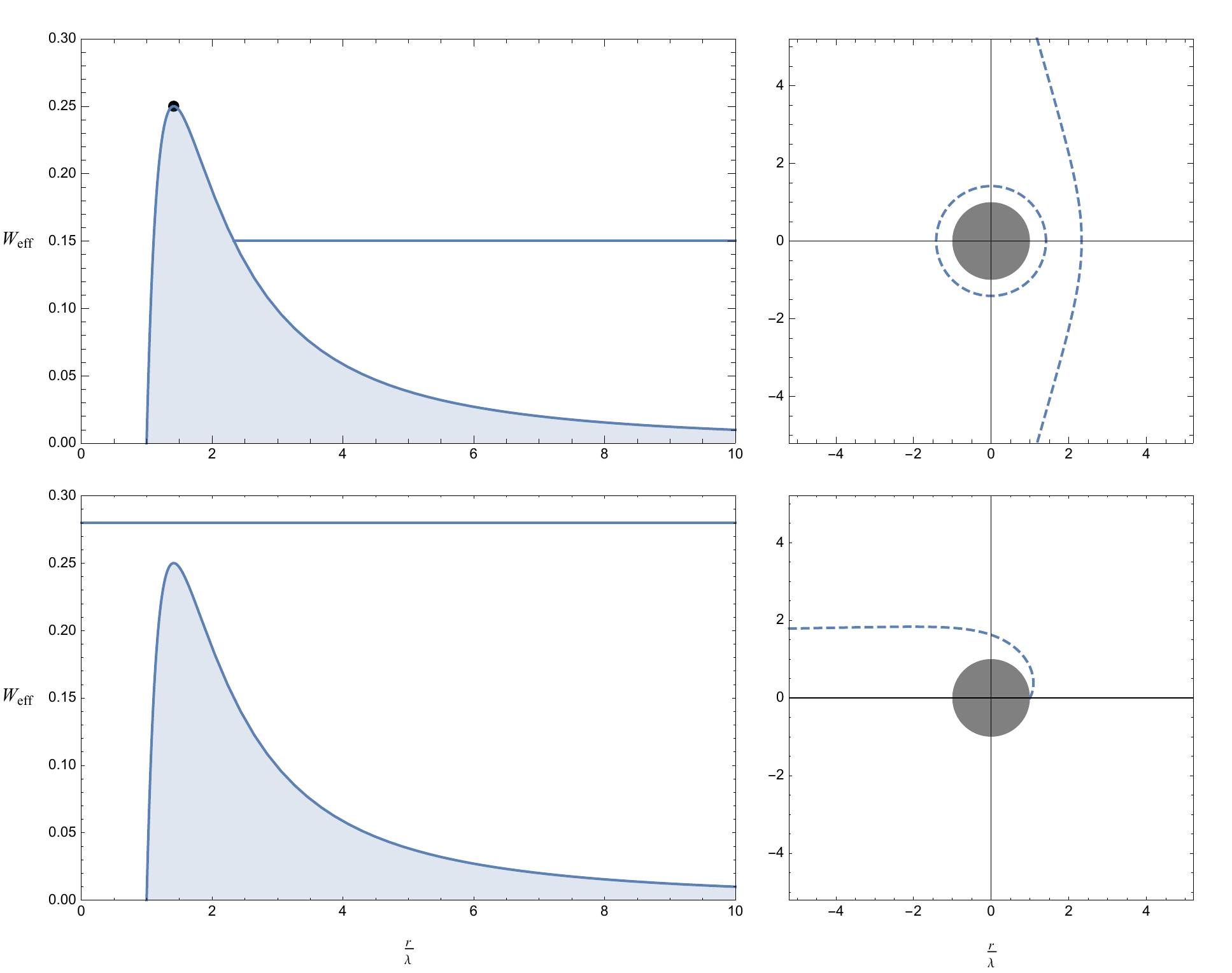}
	\caption{The picture shows three tracks corresponding to different $b$ values. The relationship between potential $W_{\mathrm{eff}}$ and $1/b^2$ is shown in the left column. The effective potential of the acoustic black hole and the behavior of the sound wave passing through the acoustic black hole: when the beam of sound wave energy is equal to the maximum value of the effective potential of the acoustic black hole, it is a critical state; when the beam of sound wave energy is less than the height of the barrier, the sound wave does not fall into the acoustic black hole; when the beam of sound wave energy is greater than the barrier height, the sound wave will fall into the black hole.}
	\label{fig:imgphononorbits}
\end{figure}

\begin{figure}[htbp]
	\centering
	\includegraphics[width=0.6\linewidth]{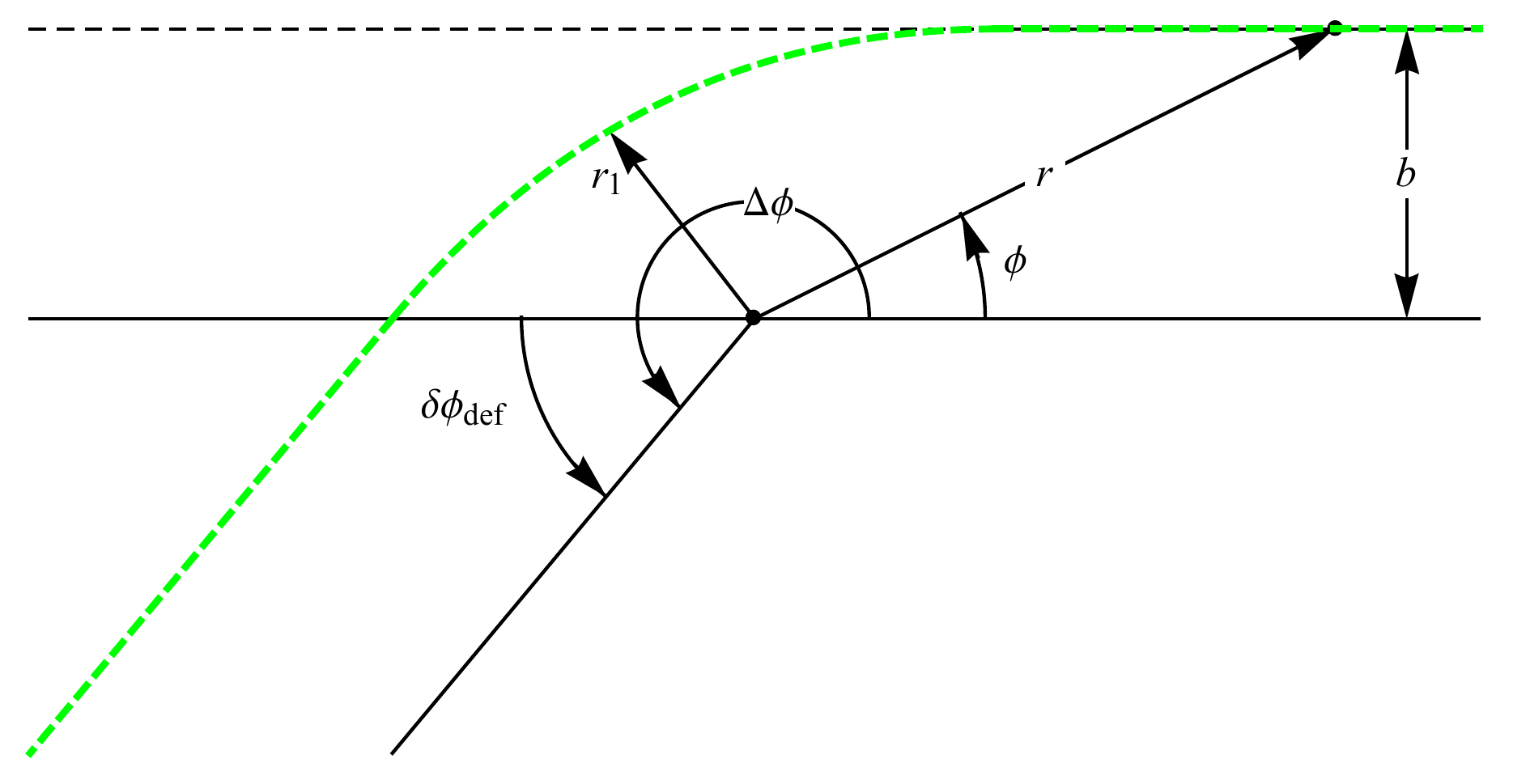}
	\caption{In this figure, $b$ is the linear distance from the black hole to the incident direction of the sound wave. A beam of sound wave coming from infinite distances pass through the vicinity of acoustic black holes and deflect in a direction of $\delta\phi_{\mathrm{def}}$.}
	\label{fig:imgdeflection}
\end{figure}

The calculation of sound wave orbits in the acoustic geometry is analogous to the calculation of vortices orbits, but with some important differences. For vortices, the unstable critical orbit does not necessarily exist, and the critical radius of the orbit is related to angular momentum, while the unstable critical radius of the acoustic wave is fixed and smaller. The world line of sound waves can be described by the coordinates $x^\alpha$ as function of  a family of affine parameters $\chi$. The null vector $u^\alpha=dx^\alpha/d\chi$ is tangent to the world line. Since the acoustic metric  (\ref{flat}) is independent of $t$ and $\phi$, the quantities
\begin{equation}\label{key12}
\epsilon=(1-v^2)\frac{dt}{d\chi},
\end{equation}
\begin{equation}\label{key13}
l=r^2\sin^2\theta\frac{d\phi}{d\chi},
\end{equation}
are conserved along sound wave orbits. If the normalization of $\chi$ is chosen so that $\mathbf{u}$ coincides with the momentum $\mathbf{p}$ of a beam of sound wave moving along the null geodesic, then $\epsilon$ and $l$ are the sound wave's energy and angular momentum at infinity. Since $\mathbf{u}\cdot\mathbf{u}=g_{\alpha \beta}\frac{dx^\alpha}{d\chi}\frac{dx^\beta}{d\chi}=0$, considering the equatorial plane condition  $\theta=\pi/2$, we have
\begin{equation}\label{key14}
-(1-v^2)(u^t)^2+(1-v^2)^{-1}(u^r)^2+r^2(u^\phi)^2=0.
\end{equation}
Utilizing equation (\ref{key12}) and equation (\ref{key13}), we have
\begin{equation}\label{key15}
-(1-v^2)^{-1}\epsilon^2+(1-v^2)^{-1}\bigg(\frac{dr}{d\chi}\bigg)^2+\frac{l^2}{r^2}=0.
\end{equation}
Multiplied by $(1-v^2)/l^2$, it can be written as
\begin{equation}\label{key16}
\frac{1}{b^2}=\frac{1}{l^2}(\frac{dr}{d\chi})^2+W_{\mathrm{eff}},
\end{equation}
therein
\begin{equation}\label{key17}
b^2=l^2/\epsilon^2,
\end{equation}
and
\be
W_{\mathrm{eff}}=\frac{1}{r^2}(1-v^2).\\\label{key18}
\ee
In the equatorial plane,
\be
x=r \cos{\phi},~~~ y=r \sin{\phi},
\ee
suppose a beam of sound wave moves parallel to the $x$-axis, and the distance from the $x$-axis is $d$, as shown in Figure \ref{fig:imgdistance}. At a distance from the source of an acoustic black hole, the sound wave moves in a straight line. For $r\gg\lambda$, the quantity $b$ is
\be\label{key19}
b=\bigg|\frac{l}{\epsilon}\bigg|=\frac{r^2}{1-v^2}\frac{d\phi}{dt}\approx r^2 \frac{d\phi}{dt}.
\ee
For very large $r$ there are $\phi\approx d/r$, and $dr/dt\approx-1$, it becomes
\be\label{pd20}
\frac{d\phi}{dt}=\frac{d\phi}{dr}\frac{dr}{dt}=\frac{d}{r^2}.
\ee
Therefore
\be
b=d.
\ee
This shows that the constant $b$ is a parameter of sound waves ray reaching infinity, and is defined to be positive.
Figure \ref{fig:imgphononorbits} shows $W_{\mathrm{eff}}$ as a function of $r$ on the left. It goes to zero at large $r$, and it has a maximum at $r=\sqrt{2}\lambda$. If $b = 2 \lambda$, the circular orbit of the sound wave at the maximum $r = \sqrt{2}\lambda$. The maximum is
\be
W_{\mathrm{eff}}(\sqrt{2}\lambda)=\frac{1}{4\lambda^2}.
\ee
If $b^2=4\lambda^2$,  it can be seen from formula \eqref{key16} that the circular orbit of the sound wave is possible at radius $r =\sqrt{2}\lambda$ in the first pair of Figure \ref{fig:imgphononorbits}. However, the circular orbit is unstable. A small perturbation in $b$ will result in orbit far away from the maximum. A stable sound wave circular orbit is impossible around the two-dimensional acoustic black hole in our model.
The qualitative characteristics of other sound wave orbits depend on whether $1/b^2$ is greater  than the maximum value of $W_{\mathrm{eff}}$, as shown in Figure \ref{fig:imgphononorbits}. First of all, consider that orbits start at infinity, if $1/b^2 < 1/4 \lambda^2$, the orbit will have a turning point and returns to infinity again,also as shown in the first line of Figure \ref{fig:imgphononorbits}. We'll discuss the size of the deflection angle in detail in a minute. If $1/b^2 > 1/4 \lambda^2$, the sound wave will fall all the way to the origin and be captured, as in the second line of Figure \ref{fig:imgphononorbits}.
The angle of interest is the deflection angle $\delta\phi_{\mathrm{def}}$, defined as in Figure \ref{fig:imgdeflection}. This angle shows a property of the shape of the sound wave orbit. This shape of a beam of sound wave can be calculated in the same way as the shape of a vortex orbit. Solve \eqref{key13} for $d\phi/d\chi$, solve \eqref{key16} for $dr/d\chi$, divide the second into the first, and then predigest using \eqref{key17} and \eqref{key18} to find
\be\label{key20}
\frac{d\phi}{dr}=\pm\frac{1}{r^2}\bigg(\frac{1}{b^2}-W_{\mathrm{eff}}\bigg)^{-\frac{1}{2}}.
\ee
The plus or minus sign indicates the direction of the orbit. When the sound wave comes in from infinity and comes back again, the total sweep angle $\mathrm{\Delta}\phi$ is twice the sweep angle from the turning point $r=r_1$ to infinity. Therefore,
\be\label{key21}
\mathrm{\Delta}\phi=2\int_{r_1}^{\infty}\frac{dr}{r^2}\bigg(\frac{1}{b^2}-\frac{1-v^2}{r^2}\bigg)^{-\frac{1}{2}}.
\ee
The radius $r_1$ satisfies ${1}/{b^2}=(1-v^2)/{r_1^2}$, i.e., the radius where the bracket in the preceding expression vanishes. Set a new variable $w$ defined by
\begin{equation}\label{key22}
r=\frac{b}{w},
\end{equation}
the expression of $\mathrm{\Delta}\phi$ turns out to be
\begin{equation}\label{key23}
\mathrm{\Delta}\phi=2\int_{0}^{w_1}dw(1-w^2(1-v^2))^{-\frac{1}{2}},
\end{equation}
where $w_1=b/r_1$ is the value of $w$ at the turning point.
For $v^2=\lambda^2/r^2=\lambda^2w^2/b^2$, we have
\bea\label{key24}
\begin{split}
	\mathrm{\Delta}\phi&=2\int_{0}^{w_1}dw\bigg(1-w^2\bigg(1-\frac{\lambda^2}{r^2}\bigg)\bigg)^{-\frac{1}{2}}\\
	&=2\int_{0}^{w_1}dw\bigg(1-w^2\bigg(1-\frac{\lambda^2 w^2}{b^2}\bigg)\bigg)^{-\frac{1}{2}}.
\end{split}
\eea
When the value of $v$ is much less than 1, i.e., $r$ is much larger than $\lambda$ and $\lambda w/b$ is far less than 1, we can recast equation (\ref{key24}) as
\begin{equation}\label{key25}
\mathrm{\Delta}\phi=2\int_{0}^{w_1}dw\bigg(1-\frac{\lambda^2 w^2}{b^2}\bigg)^{-\frac{1}{2}}\bigg(\bigg(1-\frac{\lambda^2 w^2}{b^2}\bigg)^{-1}-w^2\bigg)^{-\frac{1}{2}}.
\end{equation}
Expand the exponential term of $(1-{\lambda^2 w^2}/{b^2})$ as a power series,
\begin{equation}\label{key26}
\mathrm{\Delta}\phi=2\int_{0}^{w_1}dw\frac{1+2{\lambda^2 w^2}/{b^2}}{(1+{\lambda^2 w^2}/{b^2}-w^2)^{\frac{1}{2}}}=\frac{\pi}{(1-{\lambda^2}/{b^2})^{\frac{3}{2}}},
\end{equation}
where $w_1$ is a root of the denominator in equation (\ref{key26}).
For small $\lambda/b$, it becomes
\begin{equation}\label{key27}
\mathrm{\Delta}\phi=\pi\bigg(1+\frac{3}{2}\frac{\lambda^2}{b^2}\bigg).
\end{equation}
Therefore, the deflection angle is
\begin{equation}\label{key28}
\delta\phi_{def}=\frac{3}{2}\frac{\lambda^2}{b^2}\pi.
\end{equation}
Reinserted the factor of $c_s$ (sound velocity), the above formula can be written as ($b$ has dimensions of length)
\begin{equation}
\delta\phi_{def}=\frac{3}{2}\frac{c_s^2\lambda^2}{b^2}\pi,
\end{equation}
for small $c_s\lambda/b$. This shows that the deflection angle of the sound wave is directly proportional to the square of the $c_s\lambda/b$.

\subsection{Time delay of sound waves}
\begin{figure}[htbp]
	\centering
	\includegraphics[width=0.3\linewidth]{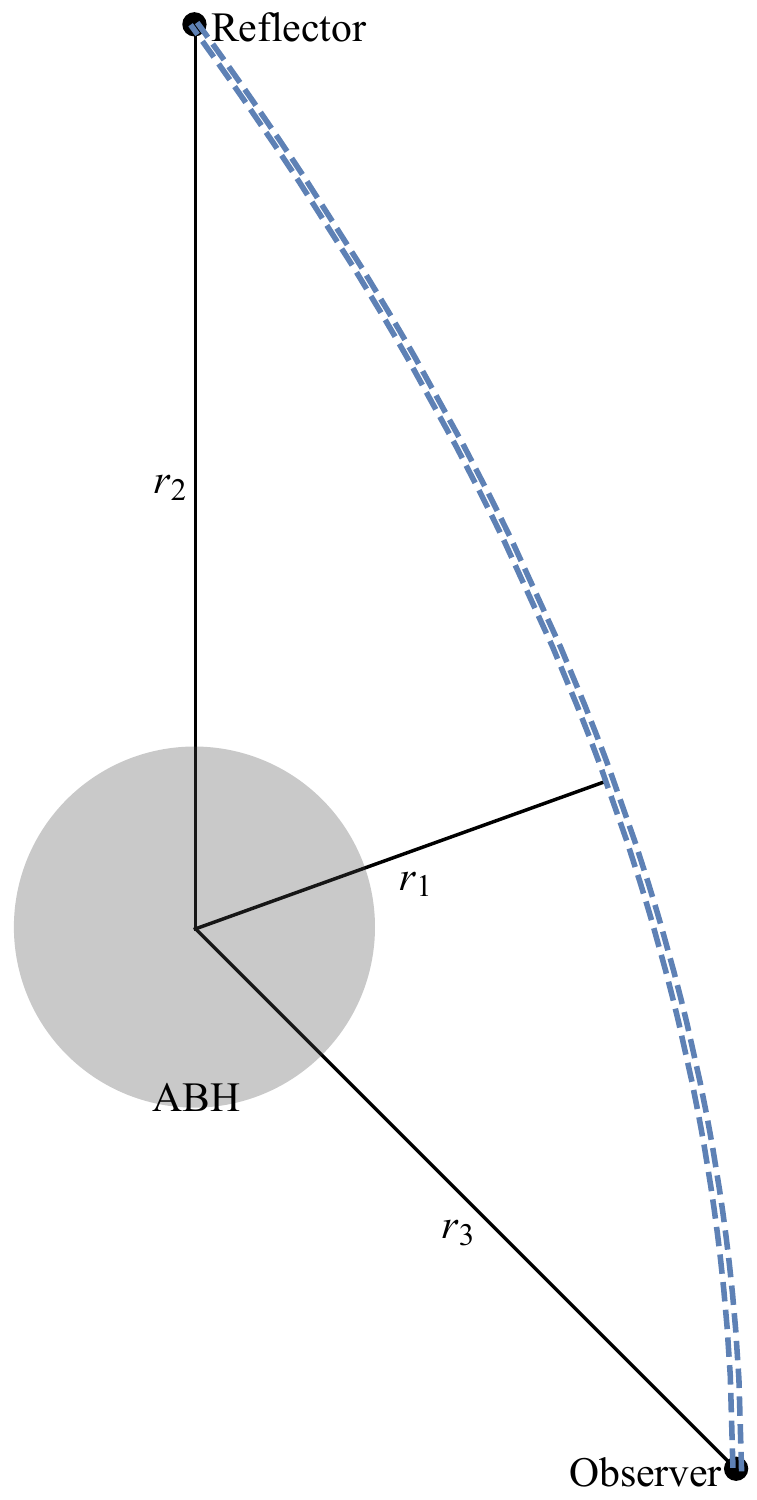}\hspace{2cm}
	\includegraphics[width=0.33\linewidth]{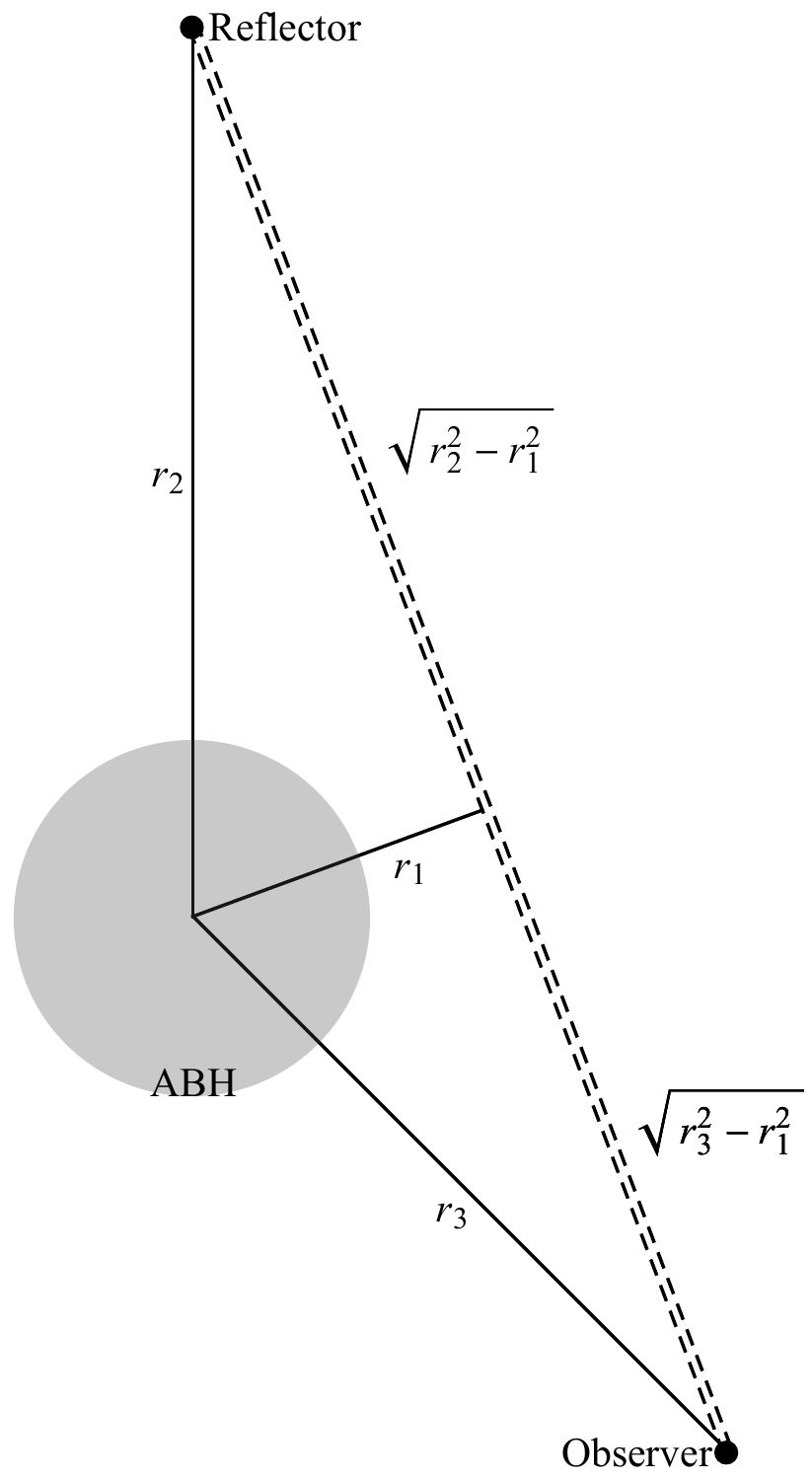}
	\caption[Sound waves bend by trajectories near acoustic black holes] {Sound waves bend by trajectories near acoustic black holes: the left figure shows the situation when the trajectory bends, while the right figure corresponds to the trajectory without bending. Obviously, when the trajectory bends, the trajectory becomes longer and the reflected signal is relatively delayed.}
	\label{fig:imgorbitcurvestraight}
\end{figure}

Since the sound wave trajectory bends near the acoustic black hole, the path of sound wave transmission at the same two points becomes longer, as shown in Figure \ref{fig:imgorbitcurvestraight}. The quantities $r_1$, $r_2$, and $r_3$ are radii of the orbits of the closest point, reflector, and observer to the acoustic black hole, respectively. Thus, the time for the sound wave to travel back and forth between the observer and the reflector will  increase. According to equation  (\ref{key12}) for $dt/d\chi$ and equation  (\ref{key16}) for $dr/d\chi$, we can obtain
\begin{equation}\label{key29}
\frac{dt}{dr}=\pm\frac{1}{b}\bigg(1-\frac{\lambda^2}{r^2}\bigg)^{-1}\bigg(\frac{1}{b^2}-W_{\mathrm{eff}}\bigg)^{-\frac{1}{2}},
\end{equation}
where the positive sign is suitable for increasing radius, and the negative sign is suitable for decreasing radius. From the left of Figure \ref{fig:imgorbitcurvestraight}, we can see that the total time for sound wave travel forth and back between observer and reflector is
\begin{equation}\label{key30}
\mathrm{\Delta} t_{total}=2t(r_1,r_2)+2t(r_1,r_3),
\end{equation}
where the time required for a beam of sound wave  moving from a radius of $r_1$ to a radius of $r$ is
\begin{equation}\label{key31}
t(r_1,r)=\int_{r_1}^{r}dr\frac{1}{b}\bigg(1-\frac{\lambda^2}{r^2}\bigg)^{-1}\bigg(\frac{1}{b^2}-W_{\mathrm{eff}}\bigg)^{-\frac{1}{2}},
\end{equation}
among them,
\begin{equation}\label{key32}
b^2={r_1^2}\bigg(1-\frac{\lambda^2}{r_1^2}\bigg)^{-1},W_{\mathrm{eff}}=\frac{1}{r^2}\bigg(1-\frac{\lambda^2}{r^2}\bigg).
\end{equation}
We consider $\lambda/r$ as a small quantity, simplify  and approximate the above formula, then
\bea\label{key33}
\begin{split}
	t(r_1,r)&=\int_{r_1}^{r}dr\bigg(1-\frac{\lambda^2}{r^2}\bigg)^{-\frac{3}{2}}\bigg(1-\frac{r_1^2}{r^2}\bigg)^{-\frac{1}{2}}\\
	&\approx\int_{r_1}^{r}dr\bigg(1+\frac{3\lambda^2}{2r^2}\bigg)\bigg(1-\frac{r_1^2}{r^2}\bigg)^{-\frac{1}{2}}.
\end{split}
\eea
The integral can be evaluated as
\begin{equation}\label{key35}
t(r_1,r)=\sqrt{r^2-r_1^2}+\frac{3\lambda^2}{2r_1}\bigg(\frac{\pi}{2}-\arctan {\frac{r_1}{\sqrt{r^2-r_1^2}}}\bigg).
\end{equation}
Thus
\begin{equation}
\mathrm{\Delta} t_{total}=\sqrt{r^2-r_1^2}+\frac{3\lambda^2}{2r_1}\bigg(\frac{\pi}{2}-\arctan {\frac{r_1}{\sqrt{r^2-r_1^2}}}\bigg).
\end{equation}
Then the radar echo delay is
\begin{equation}\label{key36}
\mathrm{\Delta} t_{excess}=\mathrm{\Delta} t_{total}-2\sqrt{r_2^2-r_1^2}-2\sqrt{r_3^2-r_1^2}.
\end{equation}
Considering $r_1/r_2\ll1,r_1/r_3\ll1$ and sound velocity $c_s$, we can recast (\ref{key36}) as
\begin{equation}\label{key37}
\mathrm{\Delta} t_{excess}\approx\frac{3c_s^2\lambda^2}{r_1}\bigg(\pi-\frac{r_1}{r_2}-\frac{r_1}{r_3}\bigg)=3c_s^2\lambda^2\bigg(\frac{\pi}{r_1}-\frac{1}{r_2}-\frac{1}{r_3}\bigg).
\end{equation}
We can see that the time delay is mostly related to $\lambda$. The larger the $\lambda$ is, the more time delay is. The parameter $\lambda$ corresponds to the flow velocity around the acoustic black hole, which is equivalent to the mass of the black hole. Note that $\lambda$ also corresponds to the radius of the horizon. Therefore, the time delay is closely related to the nature of the central acoustic black hole. Then, since $\pi/r_1$ is much larger than $1/r_2$ and $1/r_3$, this reflects that $r_1$ is the closest point to the acoustic black hole, which has a greater impact on the time delay effect. This can be understood as the closer to the acoustic black hole is, the more obvious the bending of the sound trajectory is, and the greater time delay effect is.

\section{Discussion and Conclusion}

In summary, the geometry outside a $2+1$-dimensional acoustic black hole encodes interesting physics.
For vortices falling into the acoustic black holes along the radial axis, the growth of the radial momentum near the horizon has the relation $p_{r}\sim e^{2\pi T t}$, signalizing the chaos behavior with the Lyapunov exponent $\varLambda_{\rm Lyapunov}=2\pi T$. In the case of the AdS-Schwarzschild background, the increase of the momentum near the horizon of acoustic black holes is $e^{2\pi T t}$.
These are in agreement with the chaos bound proposed in \cite{chaosbound}. The result is further confirmed from the numerical calculation of the radial infalling orbit of the vortex. It is a little bit  surprising that it saturates the chaos bound since that acoustic black holes can only simulate the kinetic aspects of real black holes, but it does not satisfy the Einstein equation. It would be interesting to verify this point experimentally.

Taking vortices as relativistic particles outside acoustic black holes, we calculated in detail the effective potential and found that there is no stable circular orbit for vortices.  As the radial falling orbit is further considered, there is no orbital precession.
Sound wave deflection and sound wave time delay are further investigated. The results show that the deflection angle of sound waves is inversely proportional to the square of the distance to the acoustic black hole. The time delay of sound waves passing through the vicinity of the acoustic black hole has the greatest correlation with the distance $r_2$ when the sound wave trajectory is closest to the acoustic black hole. It is inversely proportional to the initial distance $r_1$ and the final distance $r_3$ and $r_1$. That is to say, the greater the $r$ is, the less the effect on the time delay will be.

\section*{Acknowledgements}
We would like to thank Mikio Nakahara, Sang-Jin Sin, Yu-Qi Lei, and Hai-Ming Yuan for helpful discussions. This work is partly supported by NSFC (No.11875184).

\begin{appendix}
\section*{Appendix: Acoustic redshift}

\begin{figure}[htbp]
	\centering
	\includegraphics[width=0.7\linewidth]{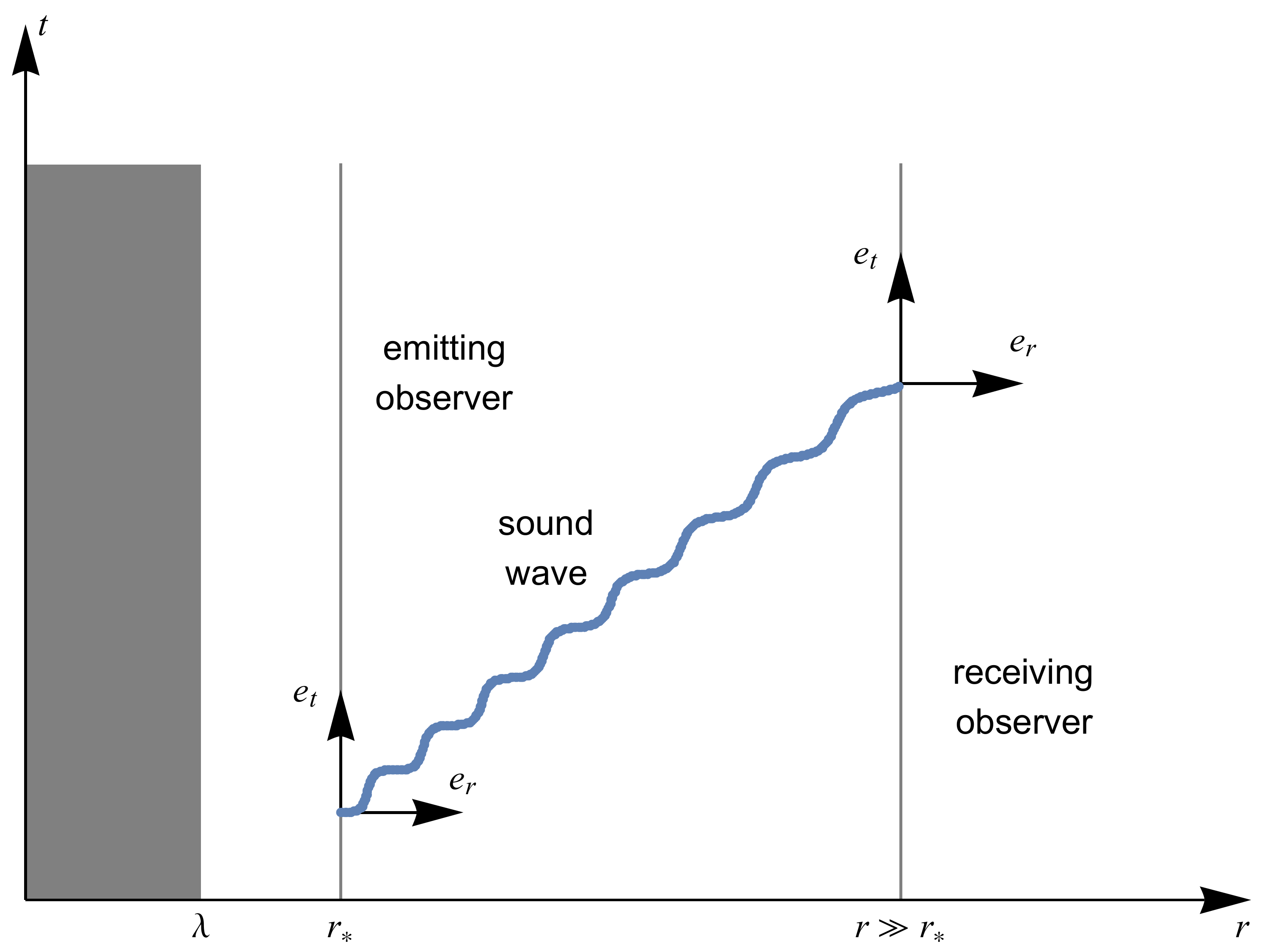}
	\caption{A sound wave traveling from $r_*$ to infinity, and the sound frequency measured by the receiver is lower than that of the sender.}
	\label{fig:imgsoundray}
\end{figure}

Let us consider an observer who emits a sound signal from a fixed radius $r_*$ near the acoustic black hole. The original signal has frequency $\omega_*$ as measured by this stationary observer. The sound signal is measured by another stationary observer(see Figure \ref{fig:imgsoundray}), where its frequency $\omega_{\infty }$ received by an observer at infinity is less than  $\omega_*$.
The sound wave energy related to the viewer's four-velocity $\mathbf{u}_{\mathrm{obs}}$ is \cite{Hartle}
\be
E=-\mathbf{p}\cdot\mathbf{u}_{\mathrm{obs}}.
\ee
Since the energy of a beam of sound wave is $E = \hbar  \omega$, $\hbar  \omega=-\mathbf{p}\cdot\mathbf{u}_{\mathrm{obs}}$ given the relationship between the observer's measured frequency and the four-velocity $\mathbf{u}_{\mathrm{obs}}$. For a stationary observer, the spatial component $u^j_{obs}$ of the four-velocity is zero ($j=r,\phi$). The time component $u^t_{obs}(r)$ at radius $r$ for a stationary observer depend on the normalization condition
\bea
\mathbf{u}_{\mathrm{obs}}( r)  \cdot \mathbf{u}_{\mathrm{obs}}( r) = g_{\alpha \beta } {u}_{\mathrm{obs}}^\alpha( r)  {u}_{\mathrm{obs}}^\beta( r)  = -1.
\eea
For static black hole, $\mathbf{u}_{\mathrm{obs}}^i(r) =0$ and this means
\be
g_{tt}(r) [u_{\mathrm{obs}}^t (r)]^2=-1.
\ee
In other words,
\be
u_{\mathrm{obs}}^t (r)=(1-v^2)^{-\frac{1}{2}}.%=(1-\frac{\lambda^2}{r^2})^{-\frac{1}{2}}.
\ee
Thus, the time-component of the four-velocity is given by
\bea
u_{\mathrm{obs}}^t(r) &=& [(1-v^2)^{-\frac{1}{2}}, 0, 0, 0] = (1-v^2)^{-\frac{1}{2}}\xi^\alpha,
\eea
where $\bm{\xi}$ is the Killing vector $\xi^\alpha=(1,0,0,0)$ corresponding to the fact that the metric is time-independent. Therefore, for a stationary observer at radius $r$, one has
\be\label{ars1}
\mathbf{u}_{\mathrm{obs}}(r) = (1-v^2)^{-\frac{1}{2}}\bm\xi.
\ee
Using  $\hbar  \omega=-\mathbf{p}\cdot\mathbf{u}_{\mathrm{obs}}$ in equation \eqref{ars1}, the frequencies of the sound wave measured by the stationary observer at radius $r_*$ and infinite radius $r_{\infty}$ are
\bea
\hbar  \omega_*  &=& (1-v_*^2)^{-\frac{1}{2}}(-\bm{\xi} \cdot \mathbf{p})_{r_*},\\
\hbar  \omega_{\infty} &=& (1-v_{\infty}^2)^{-\frac{1}{2}}(-\bm{\xi} \cdot\mathbf{p})_{r_{\infty}}.
\eea
The quantity $\bm{\xi} \cdot\mathbf{p}$ is conserved along the sound wave's geodesic. That is, $(-\bm{\xi} \cdot \mathbf{p})_{r_*}=(-\bm{\xi} \cdot\mathbf{p})_{r_{\infty}}$. Therefore, the relationship between the frequencies is
\be\label{ars2}
\omega_{\infty}=\omega_{*}\bigg(\frac{1-v_*^2}{1-v_{\infty}^2}\bigg)^{\frac{1}{2}}=\omega_{*}\bigg(\frac{1-\frac{\lambda^2}{r_*^2}}{1-\frac{\lambda^2}{r_{\infty}^2}}\bigg)^{\frac{1}{2}}=\omega_{*}\bigg(1-\frac{\lambda^2}{r_*^2}\bigg)^{\frac{1}{2}}.
\ee
In the $r\rightarrow \infty$ limit, the frequency is less than the frequency at $r_*$ by a factor $(1-\lambda^2/r_*^2)^{1/2}$. The sound wave should be influenced by the ``gravitational" redshift of the acoustic metric. If the position $r_*$ of the signal sender approaches the acoustic horizon infinitely, the frequency observed at infinity will be infinitely low.
\end{appendix}

\renewcommand{\thesection}{\Alph{section}}
\setcounter{section}{0}

\end{document}